\documentclass[letter]{aa}     % for the letters
\usepackage{graphicx}
\usepackage{txfonts}
\usepackage{lipsum}
\usepackage{subcaption}         % necessary for continued figures, example in section 3
\usepackage{lscape}             % to rotate a single page table, example in appendix.
\usepackage{placeins}           % useful with \FloatBarrier, to keep 
\def\ha{H$\alpha$}

\def\kms{\relax \ifmmode {\,\rm km\,s}^{-1}\else \,km\,s$^{-1}$\fi}

\makeatletter
\def\@LN@true{}
\def\@LN@false{}
\let\linenumbers\relax
\let\nolinenumbers\relax
\makeatother

\begin{document}

   \title{Newborn jet in the symbiotic system \hbox{R Aquarii}}

   %\subtitle{Subtitle}

    \author{T. Liimets % \orcid{0000-0003-2196-9091}
          \inst{1}   
                \and
        D. P. K. Banerjee % \orcid? 
        \inst{2}
         \and
        M. Santander-Garc\'ia % \orcid{0000-0002-7338-0986}
          \inst{3}
        \and
        J. Alcolea % \orcid{????}
           \inst{3}
        \and
            S.~B.~Howell % \orcid{0000-0002-2532-2853}
           \inst{4}
         \and
           U. Munari\inst{5} % 0000-0001-6805-9664
          \and 
         B. Deshev\inst{1} 
         \and
         C. E. Woodward\inst{6} % 0000-0001-6567-627X 
        \and 
            A. Evans\inst{7} % \orcid{0000-0002-3142-8953}
        \and
            E.~Furlan\inst{8} %\orcid{0000-0001-9800-6248} 
         \and 
         T. Geballe\inst{9}  % \orcid{0000-0003-2824-3875}
         \and
         R. D. Gehrz\inst{6} 
         \and
         V. Joshi\inst{2} 
         \and
         N. Scott\inst{10} 
         \and
         S. Starrfield\inst{11} % \orcid{0000-0002-1359-6312}
            } 
   
% inst1   
\institute{Tartu Observatory, University of Tartu, Observatooriumi 1, T\~oravere 61602, Estonia.  
\email{tiina.liimets@ut.ee}
\and 
% inst2
Physical Research Laboratory, Ahmedabad, Gujarat 380009, India
\and
% inst3
Observatorio Astron\'omico Nacional (OAN-IGN), Alfonso XII, 3, 28014, Madrid, Spain 
\and
% inst4 
NASA Ames Research Center, Moffett Field, CA 94035, USA 
\and
% inst5
INAF National Institute of Astrophysics, Astronomical Observatory of Padova, 36012 Asiago (VI), Italy
\and
%inst6 Chick
MN Institute for Astrophysics, 116 Church Street, SE University of Minnesota, Minneapolis, MN 55455, USA 
%innesota Institute for Astrophysics, School of Physics and Astronomy, 116 Church Street, S.E., University of  Minnesota, Minneapolis, MN 55455, USA
\and 
% inst8 Evans
%Astrophysics Group, Keele University, Keele, Staffordshire, ST5 5BG, UK
Astrophysics Research Centre, Lennard Jones Laboratories, Keele University, Keele, Staffordshire, ST5 5BG, UK
\and
% inst7 Furlan
NASA Exoplanet Science Institute, Caltech/IPAC, Mail Code 100-22, 1200 E. California Blvd., Pasadena, CA 91125, USA 
\and
%inst9 Geballe
Gemini Observatory/NSF’s NOIRLab, 670 N. A‘ohoku Place, Hilo, HI 96720 USA
\and
%inst10 Scott
The CHARA Array of Georgia State University, Mount Wilson Observatory, Mount Wilson, CA 91203, USA
\and 
%inst11 Sumner 
Earth and Space Exploration, Arizona State University, P.O. Box 876004, Tempe, AZ, 85287-6004, USA % \email{starrfield@asu.edu}
             }

   \date{Received September 30, 20XX}

% \abstract{}{}{}{}{}
% 5 {} token are mandatory
 
  \abstract
   %{Optional, leave empty if necessary.  The heading “Context” is used when needed to give background information on the research conducted in the paper}
   {\object{R Aquarii} (R~Aqr) is a well-known symbiotic binary that has attracted renewed interest during its recent periastron passage, an event that occurs only once every $\sim$40 years. This passage marks the first to be observed with modern, state-of-the-art instruments.}
   {We investigate the inner, sub-arcsecond active region of R~Aqr during this recent periastron passage, with the goal of gaining insight into the jet-launching mechanisms at work in this system.} 
   {We analyze \ha\ speckle interferometric images obtained one month apart using Fourier techniques. These are complemented by high-resolution optical spectra in the same emission line.}
   {Our speckle imaging reveals a newborn two-sided jet orientated in the north–south direction. Its proper motion, \hbox{$66\pm19$ mas yr$^{-1}$}, % \hbox{$0.066\pm0.019$~$''$~yr$^{-1}$}, 
   confirms that it was launched around 2020 Jan 7, at the onset of the periastron passage. Further analysis of the elongated central structure reveals a knot in the southern counterpart of the jet, moving away from the binary with 
   \hbox{$\mu=27\pm17$ mas yr$^{-1}$}  % \hbox{$\mu=0.027\pm0.017$~$''$~yr$^{-1}$} 
   at a position angle of 187$^\circ$, and an ejection time around 2019 Oct 28. This interpretation is further supported by our high-resolution spectroscopic data. In addition, we update the expansion parallax distance of R~Aqr to 260 pc.}
   %{Our speckle imaging reveals a newly detected feature whose proper motion,  $0.027\pm0.017$ $''\ yr^{-1}$, indicates it was ejected at the onset of the periastron passage, on 2019 October 28. Further analysis of the elongated central structure suggests that the jet may be two-sided. This interpretation is supported by our high-resolution spectroscopic data.}
  {}
  % {Optional, leave empty if necessary.  “Conclusions” can be used to explicit the general conclusions that can be drawn from the paper.}


   \maketitle

%%%%%%%%%%%%%%%%%%%%%%%%%%%%%%%%%%%%%%%%%%%%%%%%%%%%%%%%%%%%%%
\section{Introduction}

\object{R~Aqr} is a symbiotic binary consisting of a Mira-type variable star with a pulsation period of 387 days \citep{2000A&AS..146..407B,2009A&A...495..931G} and a white dwarf (WD) companion. At a distance of $\sim$200 pc, it is one of the closest and most studied symbiotic binaries. Despite this, R~Aqr continues to exhibit a wide range of intriguing phenomena.

The system has an orbital period of 42–44 years \citep[e.g.][]{2009A&A...495..931G,2023hsa..conf..190A}, with the WD coincidentally at inferior conjunction during each periastron passage. The most recent passage has just concluded.
The periastron passage has been proposed to enhance mass transfer from the Mira variable to WD, potentially triggering jet ejections \citep{1982Natur.298..540K}. However, other studies suggest that jet activity may not be directly related to the periastron passage \citep{2009A&A...495..931G}.

\begin{figure}[h!]
\centering
\includegraphics[width=1.0\linewidth]{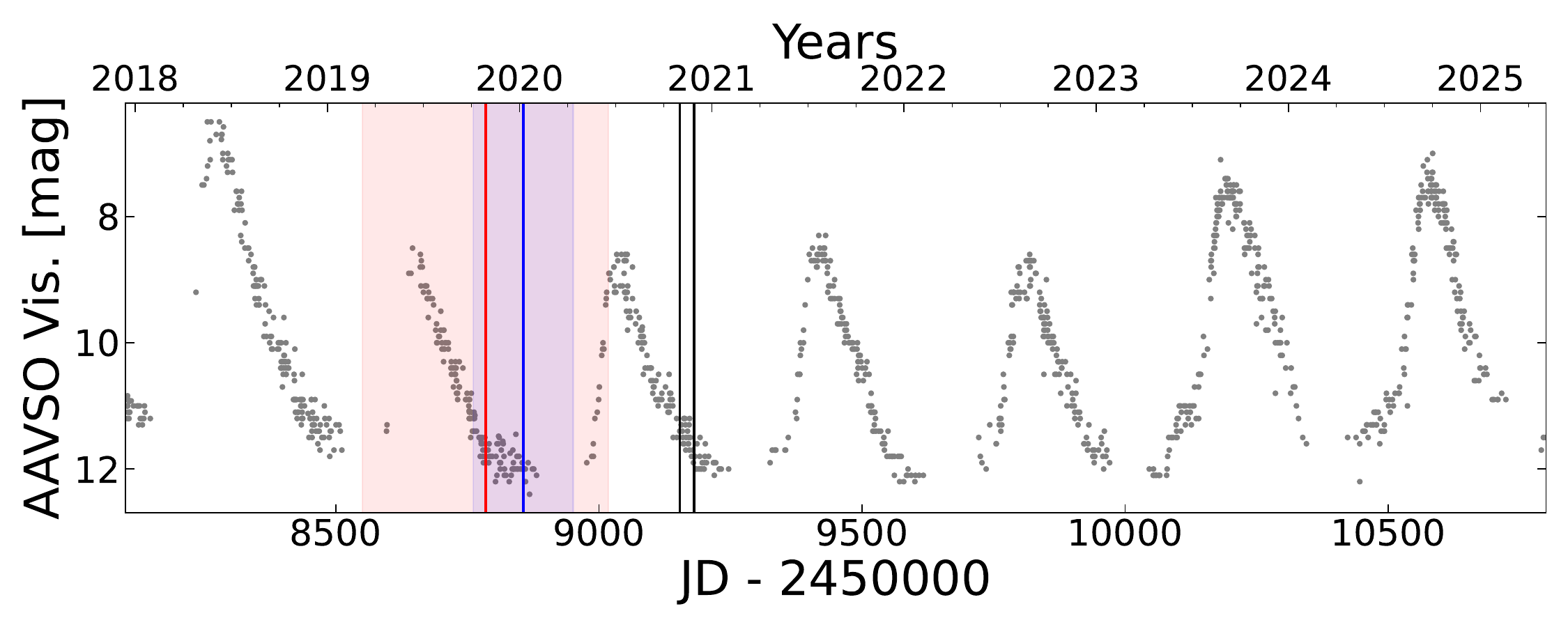}
\caption{Light curve of R~Aqr. Black lines mark Zorro observations, blue and red lines indicate the ejection of the newborn jet and the knot, respectively. Shaded regions indicate uncertainties.
}
\label{F-lc}
\end{figure}

The periastron passage is also reflected in the long-term light curve of R~Aqr, where the peak visual brightness of the Mira star pulsations diminishes by a few magnitudes every $\sim$40 years (Fig.~\ref{F-lc}). This dimming is \hbox{attributed} to increased mass transfer and enhanced dust production \citep{2025ApJ...984..128O}, obscuring the Mira variable. These dimming episodes typically last 6–7 years. The most recent event occurred between late 2018/early 2019 and mid-2023, making it slightly shorter than average. Although the WD alone is too small to eclipse the Mira star for such an extended period, the WD with its accretion disc may cause prolonged obscuration \citep{2021gacv.workE..41L}.

% plot_light_curve_palermo_ejections.py in  /home/sinope/ASTRONOOMIA/RAQR/2025_Letter/Figures/AAVSO

The binary is surrounded by a complex system of outflows (e.g. \citealt{1922PAAS....4..319L}; \citealt{1985A&A...148..274S}; \citealt{2018A&A...612A.118L} (L18); \citealt{2024MNRAS.532.2511S}), including the most recently formed component, the jet (e.g. \citealt{2017A&A...602A..53S} (S17); \citealt{2018A&A...616L...3B}). While some of these outflows extend several arcminutes from the central source, e.g. L18, this study focusses on the active, inner sub-arcsecond region during the latest periastron passage. This marks the first time that such activity around the R~Aqr’s periastron can be studied with modern, high-resolution observational facilities, 
as the previous periastron occurred about four decades ago.

%%%%%%%%%%%%%%%%%%%%%%%%%%%%%%%%%%%%%%%%%%%%%%%%%%%%%%%%%%%%%%%%

\section{Observations and data reduction}\label{S-obs}

R~Aqr and the  standard star \object{HR 8987} were observed under program GS-2020B-Q-112 using the speckle imager Zorro (\citealt{2021FrASS...8..138S}) at Gemini South in the \ha\ 656/3.22 nm and 832/40 nm filters. Observations were obtained on 2020 Oct 31 and 2020 Nov 27, respectively, under excellent seeing conditions specified in Table~\ref{T-newborn}. Forty-eight sets of exposures were taken simultaneously in both filters each night. Each of these 48 sets comprised 1000 frames, each with a duration of 0.06 seconds. The Field of View (FOV) of our observations was $2''.8\times2''.8$ with a pixel scale  of 9.758 mas~pix$^{-1}$ and 10.036 mas~pix$^{-1}$ 
%$0''.009758$ pix$^{-1}$ and  $0''.010036$ pix$^{-1}$ 
for \ha\ and 832 filters, respectively. \hbox{HR 8987} was observed before, in the middle, and after the long sequence on R~Aqr and was used in the data reduction process, which was done using Fourier techniques in a standard pipeline (e.g \citealp{2011AJ....142...19H,2012AJ....144..165H}). The 48 datasets were broken into 3 groups of 16 sets each, and these groups were reduced separately. This was done to verify the existence of any features. More details of the data reduction and interpretation techniques can be found on the Gemini-Zorro webpages\footnote{https://www.gemini.edu/instrumentation/alopeke-zorro/data-reduction}.  

Complementary spectra are reported in Appendix~\ref{A-spec}.

%%%%%%%%%%%%%%%%%%%%%%%%%%%%%%%%%%%%%%%%%%%%%%%%%%%%%%%%%%%%%%%%%%

\section{Detection of the newborn two-sided jet}\label{S-results}

\begin{figure}[t!]
\begin{center}
\includegraphics[width=1.0\linewidth]{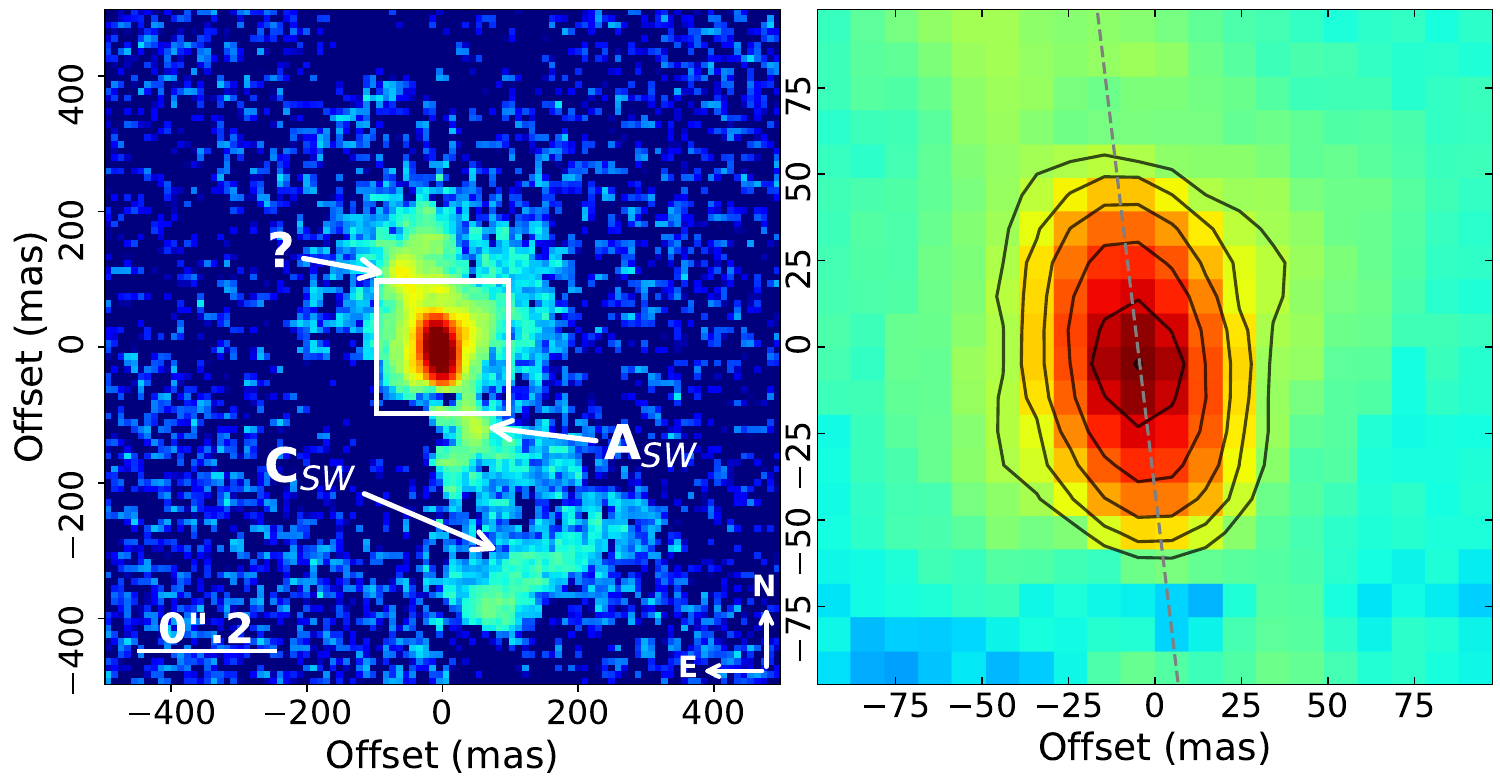}
\end{center}
\caption{\textit{Left:} Zorro \ha\ image with identified features marked. \textit{Right:} Zoom in to the central area with contours added. Dashed line shows the PA 186.9$^{\circ}$. The origin of the axes represents the geometric centre of R~Aqr. Contours are drawn at relative flux levels $F=0.03\times2^{n}$, where n=0,1,2,3,4. North is up, east is left. 
%See text for further details
}
\label{F-zorro}
\end{figure}

In Fig.~\ref{F-zorro}, we present the final combined Zorro \ha\ image from 2020 Nov 27, together with a zoom into the central area. All three groups of data collected on both dates show the same structures, therefore, we present only an image from one date in this paper. In addition, the consistent features on all groups confirm that the detected emission is real and not processing artefacts. 
Immediately, several previously known features can be noticed that provide further justification of our proper image processing,  
such as features A$_{\text{SW}}$ and C$_{\text{SW}}$, identified from SPHERE\footnote{The Spectro-Polarimetric High-contrast Exoplanet REsearch instrument at the VLT.} 2014 \ha\ image by S17. 
% marginally also B$_{\text{SW}}$ is detectable \cite{2017A&A...602A..53S}
In addition, there is a clear detection of the feature labelled as a question mark (\textbf{?}) by \cite{2021A&A...651A...4B}. All mentioned features have sustained their morphological appearances since previous observations. We will discuss these features in a forthcoming paper.
The focus of this paper is the elongated central area. 

% make_fig_2.py in /home/sinope/ASTRONOOMIA/RAQR/2025_Letter/Figures/Zorro

% profile_cuts_horisontal.py in /home/sinope/ASTRONOOMIA/RAQR/2025_Letter/From_Miguel_profile_cuts
\begin{figure}[h]
\begin{center}
\includegraphics[width=1.0\linewidth]{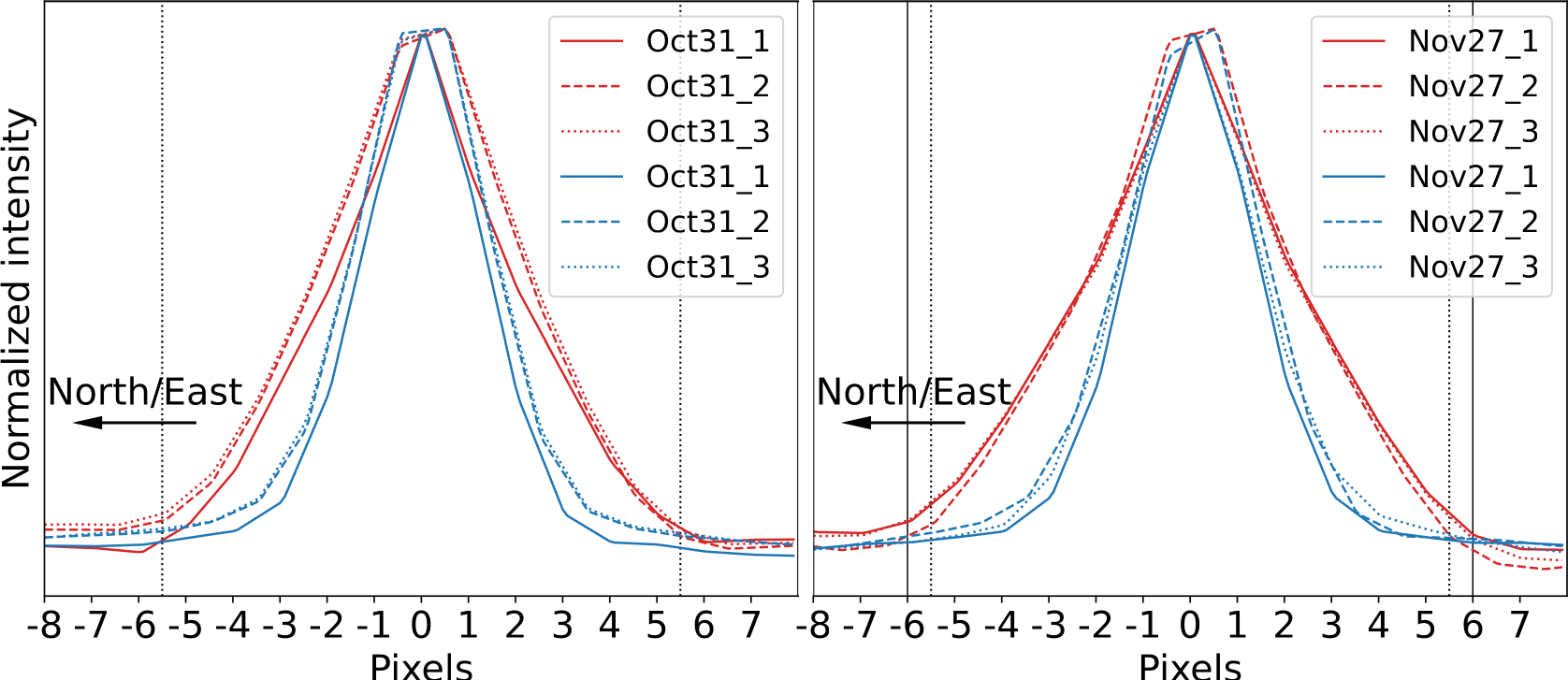}
\end{center}
\caption{Profile cuts along the major (red) and minor (blue) axes of the central elongated area for both Zorro dates. Profile cuts are centred at the Gaussian fit and flux normalized. 
}
\label{F-profilecuts}
\end{figure}

\begin{figure}[t]
\begin{center}
\includegraphics[width=1.0\linewidth]{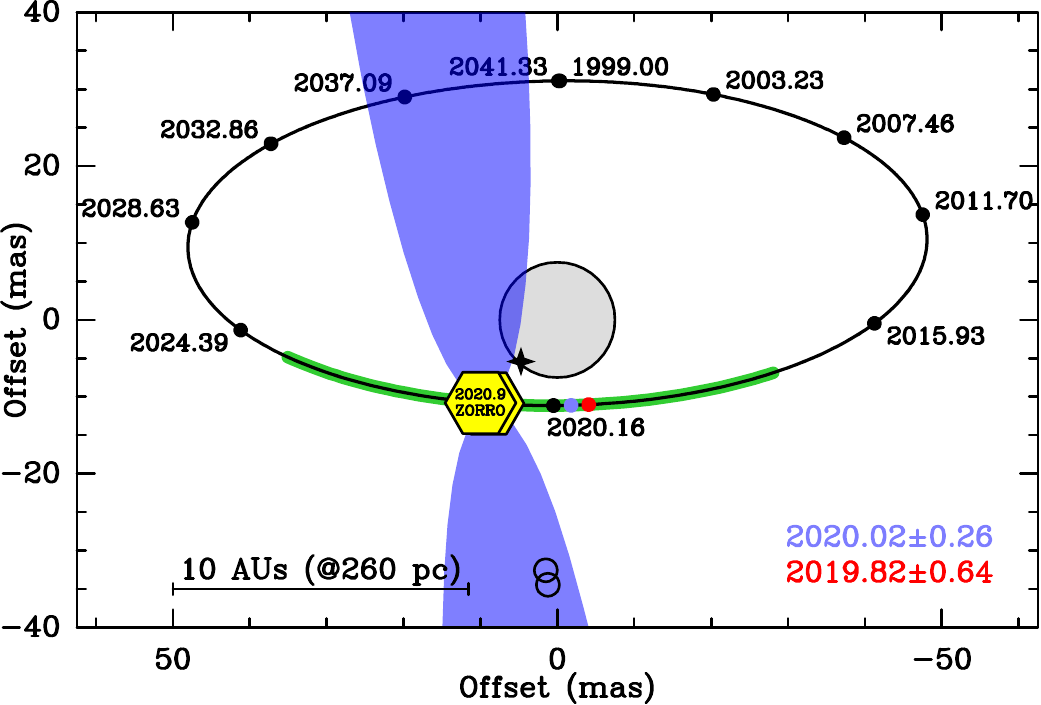}
\end{center}
\caption{Latest orbital solution of R~Aqr. Small black dots trace the WD orbit around the Mira variable (gray circle) in 0.1-phase steps, with corresponding dates in black. Zorro observing epochs (2020.82 and 2020.91) are marked by yellow hexagons. The green segment marks the recent periastron passage (2018.9–2023.3); %the blue dot indicates the two-sided jet, and the red one the knot ejection time. 
the blue dot indicates the time of two-sided jet, and the red one time of the knot ejection.
The two-sided jet is schematically drawn in blue. The black star marks the system’s geometric centre, and black circles show the knot positions at two epochs, Oct 31 and Nov 27, respectively. North is up, east is left.  %See text for further details.
}
\label{F-orbit}
\end{figure}

The central area is elongated at a position angle (PA, measured from north towards east) of $186.9\pm2.8$$^{\circ}$. To quantify this elongation,  we have plotted the profile cuts along the major and minor axes of this elongated area for all datasets on both dates (Fig.~\ref{F-profilecuts}). 
%To further verify that the central source contains an additional component, we will explore the profile cuts along and across the elongated central area.  
% From Miguel email with slight addings from Tiina, needs final text revision:
Profiles were extracted with \textsc{IRAF}\footnote{IRAF was written at the National  Optical Astronomy Observatory, which was operated by the Association of Universities for Research in Astronomy (AURA) under cooperative agreement with the National Science Foundation.} (\citealt{1986SPIE..627..733T,1993ASPC...52..173T}) using the task  \textsc{pvector} along the major and minor axes. 
%of the elongated central region. 
The orientation of the cuts is such that the left side is the northern part in profile cuts along the major axis, and the eastern part in cuts along the minor axis. The profiles are averaged, and their width is 7 pixels along the major axis and 11 pixels along the minor axis, covering the entire elongated area. 
%Profile cuts for all data groups are presented in Fig.~\ref{F-profilecuts}. 
It is clear from the figure that on both dates and in all data groups, the major axis profiles have consistently wider Full Widths at Half Maximum (FWHMs) compared to the narrower minor axis profiles, indicating an extended emitting region towards the north and south. 
The extended wings on both sides reach up to $5.5\pm0.1$ pixels from the centre on the 31st of Oct (dotted lines), while they extend up to $6.0\pm0.1$ pixels on Nov 26 (solid lines), respectively % $0''.0537$ and $0''.0585$, 
53.7 mas and 58.5 mas,
considering Zorro's pixel size (Section~\ref{S-obs}). This is a clear indication that the symmetric extended area around the bright central source has expanded between our two observing dates, which, in turn, indicates the presence of two-sided ejecta from the central source. The total extent of the two-sided jet in our first epoch is %$0''.107$, and $0''.117$ 
107 mas and 117 mas in the second. 
Proper motion, $\mu$, of the edges of the newborn two-sided jet is %$0.066\pm0.019$ $''\ yr^{-1}$, 
\hbox{$66\pm19$ mas yr$^{-1}$}, 
which translates into a tangential velocity of $v_{sky}=81\pm24$ \kms, using the formula $v_{sky}\ [\mathrm{km\ s^{-1}}] = 4.74 \cdot \mu\ [''\ \mathrm{yr^{-1}}] \cdot D \ [\mathrm{pc}]$, 
% /home/sinope/ASTRONOOMIA/RAQR/2025_Letter/Analyses/newborn.py
%\begin{equation}\label{e-vsky}
%v_{sky}\ [km\ s^{-1}] = 4.74 \cdot \mu\ [''\ yr^{-1}] \cdot D \ [pc],
%\end{equation}
taking into account our new distance estimate of $D=260$ pc discussed in Appendix~\ref{S-par}. 
%The found value of $81$ \kms\ is in accordance with previously measured velocities in both the inner and outer jets of R~Aqr (e.g. \citealp{2004A&A...424..157M}; S17; L18). Likewise, the two-sided morphology in the north-south direction, with a small inclination towards the east and west respectively, matches the already known jet characteristics in both small and large scales (e.g. \citealt{1999ApJ...514..895H}; S17; L18; \citealt{2023ApJ...947...11H}).
The measured velocity of $81$ \kms\ and the near north-south (NS) jet morphology, are consistent with previous observations of both small, and large scale jet of R~Aqr (e.g. \citealt{1999ApJ...514..895H,2004A&A...424..157M}; S17; L18; \citealt{2023ApJ...947...11H}).
When assuming a ballistic expansion, the ejection time is 2020 Jan 7 ($2020.02\pm0.26$; JD $=2458856\pm95$), which aligns well with the onset of the periastron passage and, therefore, is consistent with previous suggestions that the periastron passage is the time when the jet is actively being launched \citep{1982Natur.298..540K}. This date is marked with a blue in Figs.~\ref{F-lc} and \ref{F-orbit}.

To further support our interpretation of two-sided jet ejection, we have examined the possibility of whether the elongated central structure could be influenced by the binary components. For this purpose, in Fig.~\ref{F-orbit}, we plot the latest orbital solution of R~Aqr by \cite{2023hsa..conf..190A}. Indeed, considering the estimated position of the WD and its accretion disc during our Zorro observations (distance of $\sim$10 mas and a PA between 132.9$^{\circ}$ and 140.5$^{\circ}$ from the Mira star, yellow hexagons), the profile cuts in the direction of the minor axes could be affected. However, as can be seen, the minor axis profiles are not blended, and their FWHM are narrower than the major axis profiles. Based on this, we conclude that our Zorro observations do not resolve the binary components, and, therefore, the positions of the binary components do not affect our results.   
% With this, we conclude that our Zorro observations do not resolve the binary components. 
%However, they do resolve the newly ejected matter from the central binary of R~Aqr.  

%1) Orbital phases with a distances of 0.1 with reference to specific years in a format YYYY.Y
%2) Our Zorro observations with the WD mark (e.g. hexagonal with a text "Zorro"): years 2020.83 and 2020.91 (possibly as one marking, very nearby dates)
%3) Possible ejection time of the newborn yet with e.g. "x": 2019.82 with a range of +- 0.64 (that is 2019.18 and 2020.46), perhaps as a shaded/colored area, that would show the possible ejection period.
%4) With some lighter color the periastron passage: ~2018.9 until ~2023.3
% From Javier

We also note that the \hbox{832 nm} filter image (Fig.~\ref{F-cont} in Appendix~\ref{A-cont}) shows a point source at the location of R~Aqr. 
This wavelength is largely dominated by the primary,
%This wavelength corresponds to pure continuum emission and is free from any contribution associated with ejected material, 
confirming that the elongated structure observed in \ha\ is related to the jet activity.   
Therefore, we conclude, we have detected a newly formed two-sided jet in the innermost region of R~Aqr. 

\subsection{Knot inside the two-sided jet}\label{S-knot}

Additionally, we performed a Fourier analysis, which consistently found a fainter object in \ha\ next to the bright source on both dates and in all 3 data groups. The results of this analysis are presented in Table~\ref{T-newborn}.
%, where the distances from the central bright source and the PAs are presented together with their averaged values. 
Our measuring errors of individual distances are 0.5 mas, which are used for the average combined error. 
% From Steve Howell about the errors: "For distances (separations) measured in the speckle images, our usual error is +/- ½ mas  (0.0005 arcsec). This is based on observations of known binaries with precise orbits, and we then compare out location to the prediction for the date. It is done by measurement of a centroid of the companion star, usually ~1.2 pixels in total size. The blobs may not be as easy to centroid, but probably very close. "
The detected object is  $27.3\pm0.9$ mas and $29.3\pm0.9$ mas 
%$0''.0273\pm0''.0009$ and $0''.0293\pm0''.0009$ 
away from the central bright source, respectively, on Oct 31 and Nov 27. According to the errors, the PA on both dates does not change and is, on average, $186.9\pm2.8^{\circ}$. 
The positions of this feature are plotted with black circles in Fig.~\ref{F-orbit}. We note, as we do not resolve the central binary in our images, the positions are plotted from the geometric centre of the binary marked with black star (we are implicitly assuming that the WD and the primary equally contribute to thehttps://kosmos.ut.ee/en/stellar-physics central \ha\ emission). % If space is tight, remove the text in parentheses. 
According to measurements made on the reconstructed images, the detected source is % approximately 5.2 mag 
fainter in \ha\ than the bright nearby source, which is the unresolved central binary.
This knot is not detected in the continuum image. 

% See  /home/sinope/ASTRONOOMIA/RAQR/2025_Letter/Analyses/Zorro_newborn_calc.ods 
\begin{table}
\small
\caption{Results of the Fourier analysis in \ha\ filter.}
\label{T-newborn}
\centering
\begin{tabular}{lcccc}
\hline\hline
Date       & Data group & Distance & PA & Seeing \\
JD        &  number & (mas)   & ($^{\circ}$ ) &  ($''$)  \\     
\hline
2020-10-31 & 1  &  29.0$\pm$0.5   & 185.5  & 0.61\\
           & 2  &  27.0$\pm$0.5   & 187.0  & 0.65\\  
           & 3  &  26.0$\pm$0.5   & 186.5  & 0.77\\
2459153.53423 & &  27.3$\pm$0.9  & 186.3$\pm$0.8& \\
\hline
2020-11-27 & 1   &  30.0$\pm$0.5 & 189.5 & 0.87\\
           & 2   &  28.0$\pm$0.5 & 184.3 & 0.88\\
           & 3   &  30.0$\pm$0.5 & 188.5 & 0.91\\
2459180.55733 & & 29.3$\pm$0.9 & 187.4$\pm$2.7& \\

\hline
 \end{tabular}
 \tablefoot{The distances and the PAs of the fainter source from the brighter object are presented together with the seeing measurements for all data groups. 
 %The fourth number in column Distance is the average value for each epoch. 
The fourth row for each epoch lists the average distance and PA.
 JD refers to the mid-point of observations for each.
 }
\end{table}

% analyses_Tiina.dat and in /home/sinope/ASTRONOOMIA/RAQR/2021_Dipankar/20211230/
% /home/sinope/ASTRONOOMIA/RAQR/2025_Letter/Analyses/Zorro_newborn_calc.ods
It is evident that the detected feature is moving away from the central binary within the recently formed two-sided jet, whose overall extent from the central source is approximately twice the distance to the feature. This suggests that the observed structure corresponds to a localised enhancement of material within the jet, which we designate as a knot.

With this in mind, we can perform some numerical calculations. In order to estimate the expansion velocity of the knot in the plane of the sky, we have averaged the three measurements obtained from different data groups for each epoch (the fourth row for each data group in Table~\ref{T-newborn} indicated with the JD). We find that, between our two dates, the knot has travelled $2.0\pm1.2$ mas   % $0''.0020\pm0.0012$ 
%\textit{\textcolor{red}{(the error would be 0.0019, if I take into account the 0.0015 and 0.0012, we should mention it but it is quite large. Perhaps it is not correct to use the STDEV of the averaged values as an error? Is there any measurement error from the Fourier analysis? Perhaps those are smaller?)}}, 
assuming that the PA does not change within our error estimates. From this, we can calculate its proper motion to be $27\pm17$ mas yr$^{-1}$ % $0.027\pm0.017$ $''\ yr^{-1}$, 
which translates into a tangential velocity of $v_{sky}=33\pm21$ \kms. 
%using the formula $v_{sky}\ [km\ s^{-1}] = 4.74 \cdot \mu\ [''\ yr^{-1}] \cdot D \ [pc]$ (in convenient units, see more from \citealt{2019arXiv191004157L}), 
% /home/sinope/ASTRONOOMIA/RAQR/2025_Letter/Analyses/newborn.py
%\begin{equation}\label{e-vsky}
%v_{sky}\ [km\ s^{-1}] = 4.74 \cdot \mu\ [''\ yr^{-1}] \cdot D \ [pc],
%\end{equation}
%where $D$ is the chosen distance to R~Aqr and $\mu$ is the proper motion of the feature. Throughout the article, we use the distance of 260 pc (see Section~\ref{S-par}). 
Likewise, with the found $v_{sky}$ value of the two-sided jet, the $33$ \kms\ is in agreement with previously measured velocities in both the inner and outer jets of R~Aqr.  
% (e.g. \citealp{2004A&A...424..157M,2017A&A...602A..53S,2018A&A...612A.118L}). 
The seemingly large uncertainties, in the above calculations, are due to the small travelled distance. % influencing the error calculations. 
Assuming that the velocity of the knot has not changed, we obtain the age of $1.01\pm0.64$ years at our first observing date. 
%we can find the age of the feature at the first epoch, using the above formula in a slightly rearranged manner: 
%$T [yr] = \frac {4.74 \cdot d  \ [''] \cdot D \ [pc]} {v_{sky}\ [km\ s^{-1}]}$,
%\begin{equation}\label{e-vsky}
% T [yr] = \frac {4.74 \cdot d  \ [''] \cdot D \ [pc]} {v_{sky}\ [km\ s^{-1}]} ,
%\end{equation}
%where $d$ is the distance of the feature from the central source at the first epoch, that is $0''.0273$. 
%Using this formula, we obtain the age of $1.01\pm0.64$ years. 
This infers the ejection date of 2019 Oct 28 (2019.82 $\pm$ 0.64; JD=2458785 $\pm$ 234), which also aligns with the start of the periastron passage. This date, together with relevant errors, is marked with red in Figs.~\ref{F-lc} and \ref{F-orbit}.

\section{Discussion and conclusions}\label{S-disc_conc}

To provide additional evidence for recent jet activity in R~Aqr, we have studied a series of high-resolution \ha\ spectra, presented in  Fig.~\ref{F-sp}. %, to see whether there is any supporting kinematic signature for the jet detected in the Zorro data. 
%These spectra are not flux-calibrated, hence only comparisons of line profile shapes and radial velocities (RVs) are possible. 
All radial velocities (RVs) have been corrected for the average local standard of rest systemic velocity of R~Aqr ($-24.9$\kms, \citealp{2009A&A...495..931G}), such that the zero point in the plots corresponds to the systemic velocity. 
Fig.~\ref{F-sp} shows that the \ha\ emission line in 2012 and 2013 exhibits a single narrow profile centred near the systemic velocity. By the end of 2013, the line peak shifts slightly blueward, and by 2019, the profile becomes increasingly asymmetric toward redder wavelengths. These changes culminate in a clearly double-peaked profile by the late 2024, with a range of peak velocities between $-43.5$~\kms\ and $+21.6$~\kms.
Such a double-peaked profile could indicate
% (i) emission from an accretion disc; velocities too low
(i) bipolar mass-loss activity, such as the launching of a two-sided jet, and 
(ii) contributions from the orbital motion of the central binary.
%In the case of R~Aqr, it is likely that all three mechanisms contribute to the observed line profile. However, disentangling their relative roles is beyond the scope of this paper, therefore, we limit ourselves to a qualitative analysis. 
%%A more detailed analysis of these spectra, as also several low-resolution spectra, is planned for the future \textcolor{red}{(for Ulisse to confirm)}. 
Over the twelve years of our spectral observations, the orbital RV amplitude of the central binary does not exceed 4~\kms \citep{2009A&A...495..931G}, 
%with an average value around $-25$\kms,  
therefore, this possibility is excluded. %which has a range from $-21.6$\kms\ to $-30.7$\kms\ during the orbit (\citealp{2009A&A...495..931G}). 
%This variation alone cannot account for the significant red- and blue-shifted peaks observed in 2024. 
However, the well-established presence of a two-sided jet in the R~Aqr system (e.g. \citealt{1999ApJ...522..297H}; L18; \citealt{2024MNRAS.532.2511S}) provides a natural explanation for the range of peak velocities observed in 2024. %, spanning from $-43.5$~\kms\ to $+21.6$~\kms\ relative to the systemic velocity. 
Indeed, L18 measured peak RVs ranging from about $-40$ to $+140$~\kms\ for the large-scale jet, consistent with our new measurements.

%A remaining puzzle concerns the apparent brightness of the detected knot. Fourier analysis of the 2020 Zorro data shows that the knot was approximately 5 mag fainter than the central region. Yet in 2024, both components of the double-peaked \ha\ profile are stronger than the emission near the systemic velocity. It is well known that the brightness of R~Aqr’s large-scale jet features vary significantly over timescales of several years (L18), and even sub-arcsecond-scale structures have shown brightness variability (S17). Similar variability may have occurred in the recently ejected material, leading to its increased brightness by 2024. Additionally, the central binary itself has  become fainter due to enhanced dust production during the recent periastron passage \citep{2025ApJ...984..128O}, which obscures the emission in the optical region, as can be seen from the light curve in Fig.~\ref{F-lc}.

\cite{2024PASP..136k4504L} presents a high–spatial–resolution \ha\ image of R~Aqr obtained on 2023 July 7 with VAMPIRES\footnote{Visible Aperture-Masking Polarimetric Imager/Interferometer for Resolving Exoplanetary Signatures; \citealt{2015MNRAS.447.2894N}}. Taken 3.5 years after the proposed two-sided jet ejection, this observation provides valuable insight into the subsequent jet evolution. However, we remain cautious about their results: as noted by the authors, the WD is not detected, and the bright clump north of its expected position is attributed to new jet activity (see their Fig.~26). We find no clear reason for the WD’s absence in \ha, given that previous observations (S17) show strong emission at this wavelength, except perhaps enhanced dust production during periastron passage which would provide additional extinction in the orbital plane, but not so much along the polar directions. In any case, regardless of the WD position and the nearby northern feature, the faint north-eastern extension (upper left corner in their Fig.~26), considering the found  \hbox{$\mu=66$ mas yr$^{-1}$} could correspond to the 2020.02 ejection, 
% Some numerical calculations to support that.
while the weak southern counterpart aligns with the temporal asymmetry typically seen in R~Aqr jets (S17; L18). This asymmetry in 2023 is further supported by our spectral observations (Fig.~\ref{F-sp}).

Lastly, the ejection times of the newborn two-sided jet (2020.02) and the knot (2019.82) clearly indicate that the launching of material from the central binary is associated with the periastron passage. Within the uncertainties (see the shaded areas in Fig.~\ref{F-lc}), the ejection times are consistent with a nearly simultaneous event; however, it is possible that the slower-moving knot was ejected slightly earlier, followed shortly thereafter by the faster, two-sided outflow. This scenario is consistent with the knotty morphology and the observed range of expansion velocities of the R~Aqr jet (e.g. S17; L18). Furthermore, such a jet morphology is in agreement with hydrodynamic simulations of the jet activity in symbiotic systems (e.g. \citealt{2005A&A...429..209S}).

In conclusion, we have presented clear evidence of NS two-sided jet activity in R~Aqr during the recent periastron passage, confirming earlier suggestions that jet launching in this system occurs near periastron. This marks the first detection of the hot gas jet component so close to the central binary, revealing ejections from the accretion disc surrounding the WD. Continued monitoring will be crucial for tracing and understanding the evolution of this newborn jet, offering a rare opportunity to observe freshly ejected material in such a complex and dynamic system.

\begin{acknowledgements}

We thank the referee for their supportive report.
Observations in this paper made use of the High-Resolution Imaging instrument Zorro and were obtained under Gemini program GS-2020B-Q-112. Zorro was funded by the NASA Exoplanet Exploration Program and built at the NASA Ames Research Center by Steve B. Howell, Nic Scott, Elliott P. Horch, and Emmett Quigley. Zorro was mounted on the Gemini South telescope of the International Gemini Observatory, a program of NSF’s NOIR Lab, which is managed by the Association of Universities for Research in Astronomy (AURA) under a cooperative agreement with the National Science Foundation on behalf of the Gemini partnership: the National Science Foundation (United States), National Research Council (Canada), Agencia Nacional de Investigaci\'{o}n y Desarrollo (Chile), Ministerio de Ciencia, Tecnolog\'{i}a e Innovaci\'{o}n (Argentina), Minist\'{e}rio da Ci\^{e}ncia, Tecnologia, Inova\c{c}\~{o}es e Comunica\c{c}\~{o}es (Brazil), and Korea Astronomy and Space Science Institute (Republic of Korea).

This project has received funding from the European Union's Horizon Europe research and innovation programme under grant agreement No. 101079231 (EXOHOST), and from UK Research and Innovation (UKRI) under the UK government’s Horizon Europe funding guarantee (grant number 10051045). % Tiina

This work has been partially supported by the I+D+i PID2023-146056NB-C21 (CRISPNESS/MESON), funded by the AEI (10.13039/501100011033) of the Spanish MICIU and the European Regional Development Fund (ERDF) of the EU. % Miguel and Javier

% This was minimal if at all used, can be left out if no space
This research made use of the NASA Astrophysics Data System (ADS) and of the SIMBAD database, which is operated at CDS, Strasbourg, France. 

%Ulisse
This work has been in part supported by INAF 2023 MiniGrant Program (contract C93C23008470001 to UM.

% AAVSO
We acknowledge with thanks the variable star observations from the AAVSO International Database contributed by observers worldwide and used in this research.

% Sumner
SS acknowledges partial support from a NASA Emerging Worlds grant to ASU (80NSSC22K0361) as well as support from his ASU Regents’ Professorship monies.

RDG was supported, in part, by the United States Air Force.

\end{acknowledgements}

\bibliographystyle{aa}
\bibliography{literature}

\begin{appendix} 

% Tiina is working in here. 

\section{Complementary spectral observations}\label{A-spec}

Complementary spectral observations were collected with the 1.82 m Asiago Echelle Spectrograph between the years 2012 and 2024. Twenty spectra were collected in the region of \ha\ with a spectral resolution $R = 20 000$ and a typical exposure time of $3\times300$ seconds. The slit was $2''$ wide and aligned with the north-south (NS) direction. Given the low declination, R~Aqr was always observed close to the meridian, and in such conditions, the NS orientation is always nearly coincident with the parallactic angle. Furthermore, the NS slit orientation ensures that all features visible in our Zorro images were included in the slit.
%Ulisse: the slit was 'oriented North-South at all epochs. Given the low declination, R~Aqr was always observed close to transiting at meridian, and in such conditions the North-South orientation is coincident with the paralactic angle'. 
Asiago spectra were reduced using standard routines in \textsc{IRAF}.

\begin{figure}{b}
\begin{center}
\includegraphics[width=0.8\linewidth]{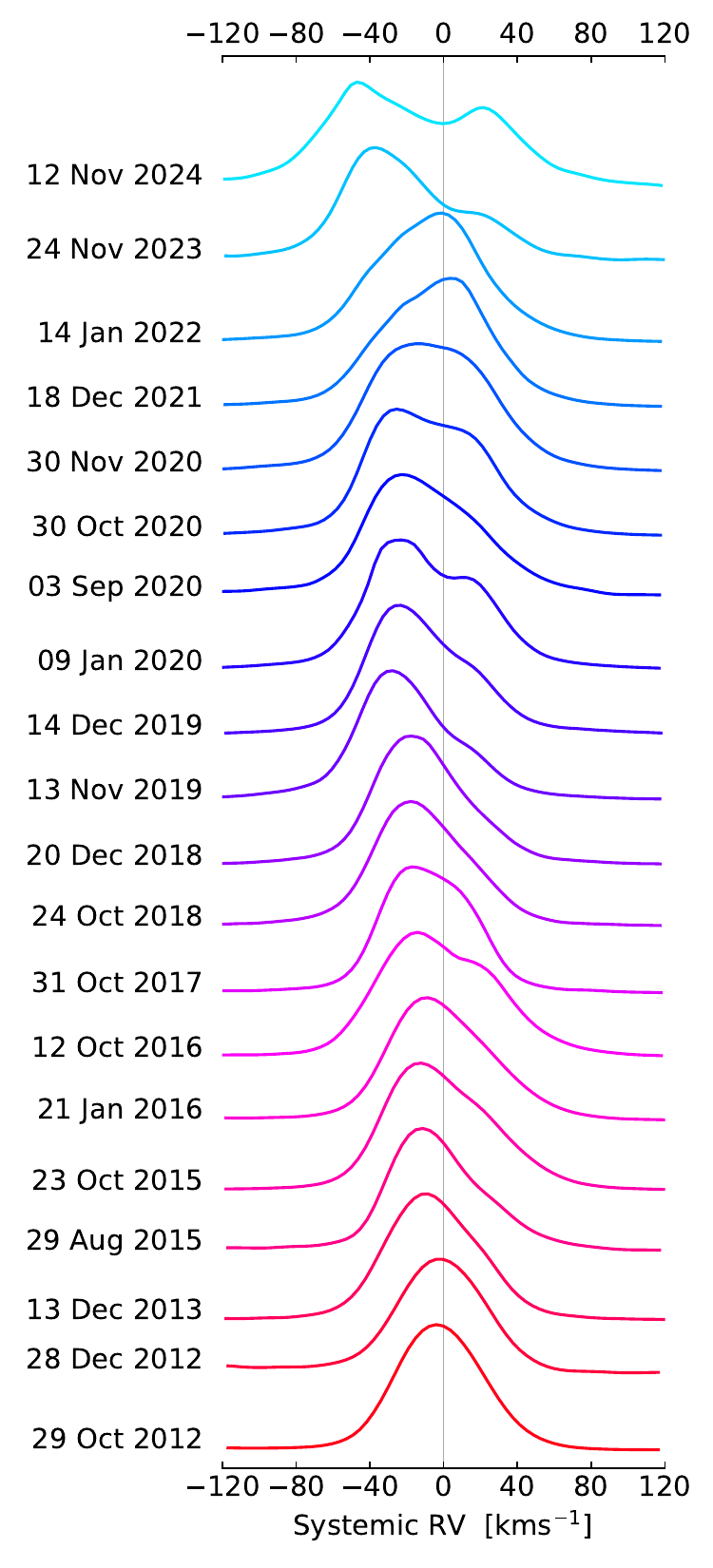}
\end{center}
\caption{\ha\ line profiles of the Asiago spectra. Zero point represents the average systemic velocity of R~Aqr, $-24.9$ \kms.}
\label{F-sp}
\end{figure}

The \ha\ profiles are shown in Fig.~\ref{F-sp}. All RVs have been corrected for the average systemic velocity of R~Aqr ($-24.9$~\kms; \citealp{2009A&A...495..931G}), so that the zero point in the plots corresponds to the systemic velocity.

% plot_figure.py in /home/sinope/ASTRONOOMIA/RAQR/2025_Letter/Figures/Asiago

\section{Revised expansion parallax}\label{S-par}

In order to have reliable values of physical sizes and tangential velocities, it is important to use an accurate distance to the object. 
Here, we re-examine the previously determined expansion parallax $178\pm18$~pc (L18). This value was 
% From Miguel
 computed using the magnification factor of the hour-glass nebula at the equator, and comparing it to the equatorial ring expansion velocity of $v_{exp}=55$ \kms\ modelled by \cite{1985A&A...148..274S}. More recent morphokinematical analysis by \cite{2024MNRAS.532.2511S} used long-slit high resolution spectra, finding a considerably larger expansion velocity for the equatorial ring of $v_{exp}=86\pm5$ \kms. This prompts a new distance determination using the magnification factor by L18, and the size ($r=45.6$ arcsec) and expansion velocity of the equatorial ring by \cite{2024MNRAS.532.2511S}. Assuming a conservative error of $2''$ for the size of the ring, and the standard deviation of the four measurements of the magnification by L18 as the uncertainty of the magnification factor, the expansion parallax distance to R~Aqr is $260\pm27$~pc. 
Although slightly larger than our previous estimate, this value is in good agreement with several other determinations based on nebular kinematics: 260~pc \citep{1944MWOAR..16....1A} and 273~pc \citep{2005A&A...435..207Y}, as well as with the period–luminosity relation distance of 250~pc \citep{2008MNRAS.386..313W}. Moreover, the revised distance aligns significantly better with the most recent orbital solution and binary parameters, leading to more plausible stellar masses. 

According to the orbital solution by \cite{2023hsa..conf..190A}, the total mas of the system is 3.3$\times$(D(pc)/333)$^3$. \cite{2014PASJ...66...38M} derived a distance of 218 pc [208--230 pc] based on SiO masers VLBI parallax, resulting in a total mass of 1.0 M$_\odot$ for the system, being both components about 0.5 M$_\odot$.
These values are unacceptably low, not to say those resulting from the previous estimate of 178 pc. \cite{2021AJ....161..147B} estimates a distance of 387 pc [343--436 pc] from GAIA DR3 results, but these measurements are not reliable according to several flags in the catalogue ($\tt ruwe$\footnote{Renormalised Unit Weight Error. ${\tt Ruwe}$ is expected to be around 1.0 for sources where the single-star model provides a good fit to the astrometric observations. A value significantly greater than 1.0 (e.g., >1.4) could indicate that the source is non-single or otherwise problematic for the astrometric solution.} and ${\tt ipd\_frac\_multi\_peak}$\footnote{Per cent of successful-IPD windows with more than one peak.}). In addition, a distance of 387 pc will result in a total mass of 5 M$_\odot$ with a primary of 3.5 M$_\odot$ and a WD mass of 1.5 (but with large errors) that are unacceptably high. On the contrary, for a distance of 260 pc, we get a total mass of 1.75  M$_\odot$, with a primary of 1  M$_\odot$ and a secondary of 0.75  M$_\odot$, which implies that the secondary had an initial mass of about 3 M$_\odot$ star. These numbers are much more plausible in view of other properties of the primary, such as its C/O abundance ratio less than one (R Aqr A is an O-rich Mira star), and the low $^{17}$O/$^{18}$O ratio of 0.05 \citep{2016ApJ...825...38H}.

%%%%%%%%%%%%%%%%%%%%
\newpage
\section{Continuum image}\label{A-cont} 

\begin{figure}[h]
\begin{center}
\includegraphics[width=1.0\linewidth]{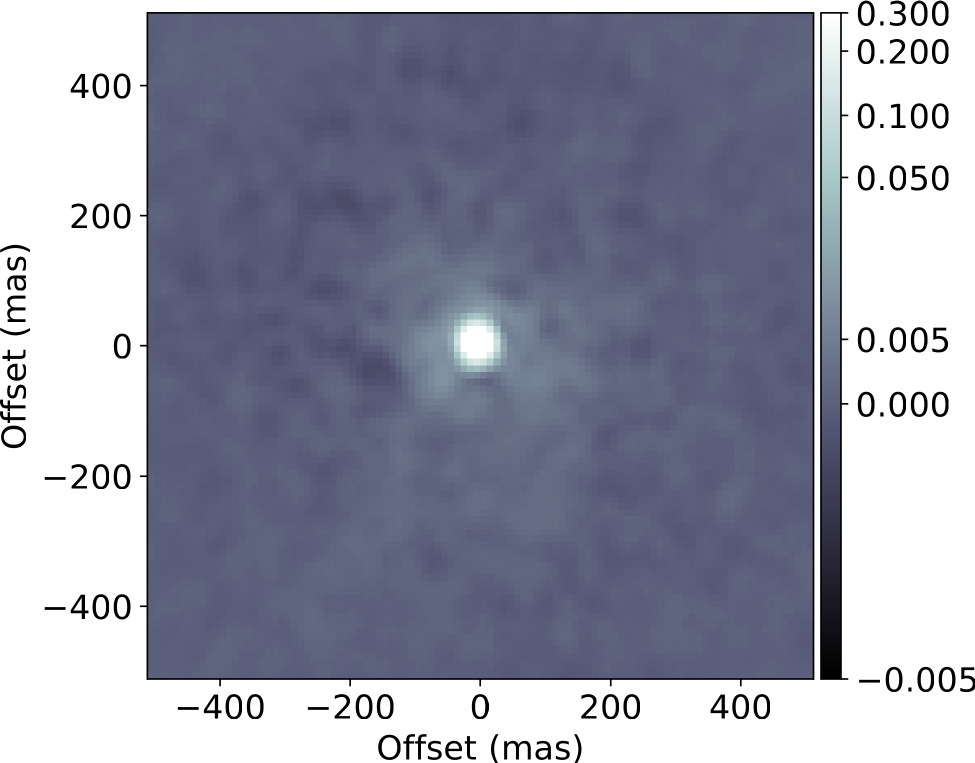}
\end{center}
\caption{Zorro image obtained with the 832 nm filter. The sky area shown is the same as in the left panel of Fig.~\ref{F-zorro}. North is up and east is left.
}
\label{F-cont}
\end{figure}

In Fig.~\ref{F-cont}, we present a representative image obtained with the 832 nm filter on 2020 Nov 27. At this wavelength, the emission is primarily dominated by the Mira variable. The central source appears roughly round in shape. 

\end{appendix}

\end{document}